\PassOptionsToPackage{obeyspaces}{url}
\documentclass[sigconf,screen]{acmart}

\usepackage[utf8]{inputenc}
\usepackage[T1]{fontenc}
\usepackage{microtype}

\usepackage{bm}
\usepackage{graphicx}
\usepackage{amsmath,amsfonts}
\usepackage{booktabs}
\usepackage{multirow}
\usepackage{makecell}
\usepackage{textcomp}
\usepackage{array}
\usepackage{textcomp}
\usepackage[font=footnotesize]{subfig}
\usepackage{multirow}
\usepackage{float}
\usepackage{xspace}
\usepackage[export]{adjustbox}
\usepackage{nicefrac}
\usepackage{colortbl}
\usepackage{tablefootnote}
\usepackage[referable]{threeparttablex}
\usepackage{comment}
\usepackage{enumitem}
\usepackage{balance}

\usepackage[linesnumbered,ruled,vlined]{algorithm2e}
\SetKwRepeat{Do}{do}{while}
\DontPrintSemicolon

\SetCommentSty{commentsty}

\usepackage{mathtools}

\DeclarePairedDelimiterX{\infdivx}[2]{(}{)}{%
	#1\;\delimsize\|\;#2%
}

\makeatletter
\@ifclassloaded{acmart}{
\def\authornotetext#1{
	\g@addto@macro\@authornotes{%
	\stepcounter{footnote}\footnotetext{#1}}%
}}{
\usepackage[usenames,svgnames]{xcolor}
\usepackage{hyperref}
\hypersetup{
    colorlinks = true,
    breaklinks = true,
    bookmarksdepth = section,
    citecolor = DarkGreen,
    linkcolor = DarkRed,
    urlcolor = DarkBlue,
}}
\@ifclassloaded{llncs}{
\usepackage[square,comma,numbers,sort&compress]{natbib}
\bibliographystyle{splncsnat}

}{
\usepackage{amsthm}

\theoremstyle{remark}

}
\makeatother

\DeclareMathOperator*{\argmax}{argmax}
\DeclareMathOperator*{\argmin}{argmin}

\DeclareMathAlphabet{\mathsfit}{\encodingdefault}{\sfdefault}{m}{sl}
\SetMathAlphabet{\mathsfit}{bold}{\encodingdefault}{\sfdefault}{bx}{n}

\usepackage[ruled,linesnumbered]{algorithm2e}

\ccsdesc[300]{Information systems~Collaborative filtering}
\ccsdesc[500]{Information systems~Recommender systems}

\copyrightyear{2023}
\acmYear{2023}
\setcopyright{acmlicensed}\acmConference[SIGIR '23]{Proceedings of the 46th International ACM SIGIR Conference on Research and Development in Information Retrieval}{July 23--27, 2023}{Taipei, Taiwan}
\acmBooktitle{Proceedings of the 46th International ACM SIGIR Conference on Research and Development in Information Retrieval (SIGIR '23), July 23--27, 2023, Taipei, Taiwan}
\acmPrice{15.00}
\acmDOI{10.1145/3539618.3591729}
\acmISBN{978-1-4503-9408-6/23/07}

\author{Jinghao Zhang}
\affiliation{
	\institution{CRIPAC, MAIS,\\Institute of Automation, Chinese Academy of Sciences}
    \institution{School of Artificial Intelligence,\\University of Chinese Academy of Sciences}
    \country{Beijing, China}
}
\email{jinghao.zhang@cripac.ia.ac.cn}

\author{Qiang Liu}
\affiliation{
	\institution{CRIPAC, MAIS,\\Institute of Automation, Chinese Academy of Sciences}
    \institution{School of Artificial Intelligence,\\University of Chinese Academy of Sciences}
    \country{Beijing, China}
}
\email{qiang.liu@nlpr.ia.ac.cn}

\author{Shu Wu}
\authornote{Corresponding authors}
\affiliation{
	\institution{CRIPAC, MAIS,\\Institute of Automation, Chinese Academy of Sciences}
    \institution{School of Artificial Intelligence,\\University of Chinese Academy of Sciences}
    \country{Beijing, China}
}
\email{shu.wu@nlpr.ia.ac.cn}

\author{Liang Wang}
\affiliation{
	\institution{CRIPAC, MAIS,\\Institute of Automation, Chinese Academy of Sciences}
    \institution{School of Artificial Intelligence,\\University of Chinese Academy of Sciences}
    \country{Beijing, China}
}
\email{wangliang@nlpr.ia.ac.cn}

\begin{document}

\newcommand{\themodel}{MODEST\xspace}

\title{Mining Stable Preferences: Adaptive Modality Decorrelation for Multimedia Recommendation}

\begin{abstract}
Multimedia content is of predominance in the modern Web era. Many recommender models have been proposed to investigate how users interact with items which are represented in diverse modalities. In real scenarios, multiple modalities reveal different aspects of item attributes and usually possess different importance to user purchase decisions. However, it is difficult for models to figure out users' true preference towards different modalities since there exists strong statistical correlation between modalities. Even worse, the strong statistical correlation might mislead models to learn the spurious preference towards inconsequential modalities. As a result, when data (modal features) distribution shifts, the learned spurious preference might not guarantee to be as effective on the inference set as on the training set. 

Given that the statistical correlation between different modalities is a major cause of this problem, we propose a novel \underline{MO}dality \underline{DE}correlating \underline{ST}able learning framework, \themodel for brevity, to learn users' stable preference. Inspired by sample re-weighting techniques, the proposed method aims to estimate a weight for each item, such that the features from different modalities in the weighted distribution are decorrelated. We adopt Hilbert Schmidt Independence Criterion (HSIC) as independence testing measure which is a kernel-based method capable of evaluating the correlation degree between two multi-dimensional and non-linear variables. Moreover, by utilizing adaptive gradient mask, we empower HSIC with the ability to measure task-relevant correlation. Overall, in the training phase, we alternately optimize (1) model parameters via minimizing weighted Bayesian Personalized Ranking (BPR) loss and (2) sample weights via minimizing  modal correlation in the weighted distribution (quantified through HSIC loss). In the inference phase, we can directly use the the trained multimedia recommendation models to make recommendations. Our method could be served as a play-and-plug module for existing multimedia recommendation backbones. Extensive experiments on four public datasets and four state-of-the-art multimedia recommendation backbones unequivocally show that our proposed method can improve the performances by a large margin.
\end{abstract}

\keywords{Multimedia Recommendation, Stable Preference, Out-of-Distribution, Stable Learning}

\maketitle
\newcommand{\inlinegraphics}[1]{(\raisebox{-.1\height}{\includegraphics[width=0.9em]{#1}})}

\section{Introduction}
\begin{figure}
	\centering
	\includegraphics[width=\linewidth]{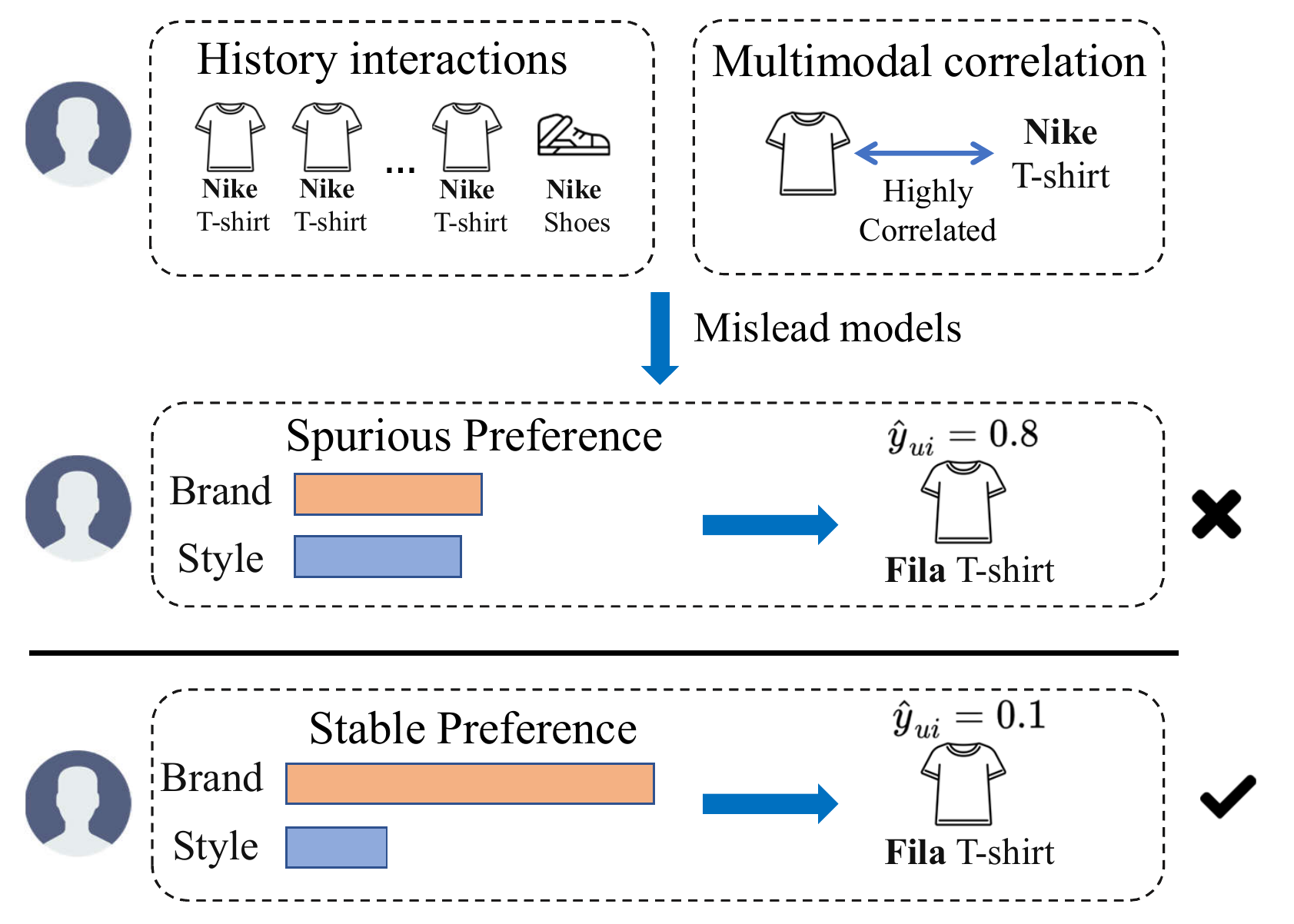}
\caption{A toy example to illustrate spurious preference in multimedia recommendation. Models might be mislead to learn spurious preference due to the strong correlation between modalities.}
\label{fig:toy}	
\end{figure}

Personalized recommender systems have become an indispensable tool to help users find relevant information from massive irrelevant contents on the Internet. Nowadays, a large portion of Internet contents are represented in multiple modalities, including images, texts, videos, etc. For example, users usually choose products based on the visual appearance and textual descriptions of products in e-commerce platform; both video contents and textual tags are important for users to find instant videos they are interested in.

Recent years have witnessed growing research interests in multimedia recommendation, which aims to incorporate multimodal content information of items into collaborative filtering (CF) framework. For example, VBPR \cite{He:2016ww} extends the framework of Bayesian Personalized Ranking (BPR) by concatenating content vectors with item embeddings. DeepStyle \cite{Liu:2017ij} derives user preferences for different styles from visual contents. ACF \cite{Chen:2017jj} introduces an item-level and component-level attention model to better understand the user preferences for item contents. Inspired by the success of graph neural networks \cite{Kipf:2017tc,Velickovic:2018we}, many attempts have been made to integrate multimodal contents into graph-based recommendation systems. For example, MMGCN \cite{Wei:2019hn} constructs modality-specific user-item interaction graphs to model user preferences specific to each modality. GRCN \cite{Wei:2020ko} refines user-item interaction graph by identifying the false-positive feedback and prunes the corresponding noisy edges in the interaction graph.

Despite their effectiveness, we argue that previous methods might be misled to learn \textbf{\textit{spurious preference}} towards inconsequential modality due to the strong statistical correlation between different modalities. As a result, the stable preference which possesses greater causal effect on the users' purchase decision might not be emphasized.  Specifically, taking Figure \ref{fig:toy} as an example, we dissect such problem in multimedia recommender systems into four parts: 
(1) \textbf{Multiple modalities reveal different aspects of item attributes and usually possess different importance to user purchase decisions.} If the user has purchased Nike T-shirts, most existing models will treat both textual description (Nike T-shirt) and visual style \inlinegraphics{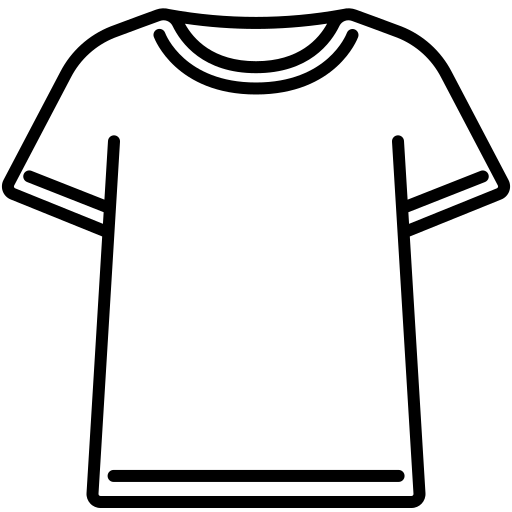} as the user's preferred attributes. However, it is likely that the user is only a fan of Nike, and will not buy a similar style T-shirts of other brands. 
(2) \textbf{There exists strong statistical correlation between different modalities.}  In the training set, the majority of Nike T-shirts have similar visual style. That is, the visual style \inlinegraphics{figures/tshirt} and textual description (Nike T-shirts) usually co-occur and thus are strongly correlated. 
(3) \textbf{Models might learn the spurious preference towards inconsequential modalities.} Since the visual style \inlinegraphics{figures/tshirt} and textual description (Nike T-shirts) are strongly correlated, models might have difficulty in identifying the user's true intent and might mistakenly suppose that the user also likes the visual style \inlinegraphics{figures/tshirt}.
(4) \textbf{The learned spurious preference might not guarantee to be as effective on testing set as on the training set, especially when data distribution shifts.} When encountering ``FILA T-shirt'' with similar visual style, the model is prone to utilize the spurious preference to make false recommendations. On the contrary, the stable preference reflects true causal effect and would not recommend FILA T-shirt to the user.

Given that the statistical correlation between different modalities is a major cause of such spurious dependency, we propose a novel \underline{MO}dality \underline{DE}correlating \underline{ST}able learning framework, \themodel for brevity. \themodel could be served as a play-and-plug module for existing multimedia recommendation backbones to guide models to focus on the modalities which can reveal users' true preference even when data distribution shifts. Inspired by sample re-weighting techniques \cite{kuang2020stable, shen2020stable, zhang2021deep, zhang2021stable}, the proposed \themodel aims to estimate a weight for each item such that the features of different modalities in weighted distribution are decorrelated. Specifically, we adopt Hilbert Schmidt Independence Criterion (HSIC) as independence testing measure \cite{gretton2005measuring, gretton2007kernel, greenfeld2020robust}, which is a kernel-based method capable of evaluating the correlation degree between two multi-dimensional and non-linear variables. Moreover, the vanilla HSIC essentially measures task-agnostic correlation, which might be sub-optimal for multimedia recommendation task since users usually pay different attention to various item attributes. To evaluate task-relevant correlation, we first define the adaptive mask of each dimension of the modal features as the normalized gradients from the final predictions to them, and calculate the HSIC between masked modal features that are highly important to multimedia recommendation task.

The overall framework of our proposed \themodel is shown in Figure \ref{fig:model}. In the training phase, we alternately optimize model parameters and sample weights. At each epoch, \themodel first fixes the sample weights and optimizes the trainable parameters of multimedia recommendation models according to the weighted BPR loss. After that, \themodel finds optimal sample weights by minimizing the task-relevant HSIC between the weighted features of different modalities while keeping the trainable parameters of multimedia recommendation models fixed. The sample weights are only used in the training phase to guide model training. In the inference phase, we can directly use the trained multimedia recommendation models to make recommendations. Please note that since different modalities of each item are correlated inherently, we do not directly change the features which might break the inherent correlation. Instead, our design can assign high weights to the items of which the correlation is weak, and thus backbone models are guided to pay more attention to these hard samples.

We conduct extensive experiments to verify the effectiveness of our proposed \themodel on four public real-world recommendation datasets. Additionally, we change the data split to simulate data distribution shifts. Experimental results demonstrate that our proposed \themodel gains significant improvements when plugged into four state-of-the-art models.  

To summarize, the main contribution of this work is fourfold.
\begin{itemize}
	\item We argue that the strong statistical correlation between different modalities might mislead models to learn statistical spurious preference towards inconsequential modalities.
	\item We propose a novel stable learning framework for multimedia recommendation to decorrelate different modalities and learn users' stable preference. We evaluate task-relevant correlation by enhancing vanilla HSIC with adaptive task-relevant mask.
	\item We perform extensive experiments on four public datasets when plugged into four backbones. The empirical results validate the effectiveness of our proposed model. 
\end{itemize}
\section{Preliminaries and Pilot Study}
\begin{figure}
	\centering
	\includegraphics[width=\linewidth]{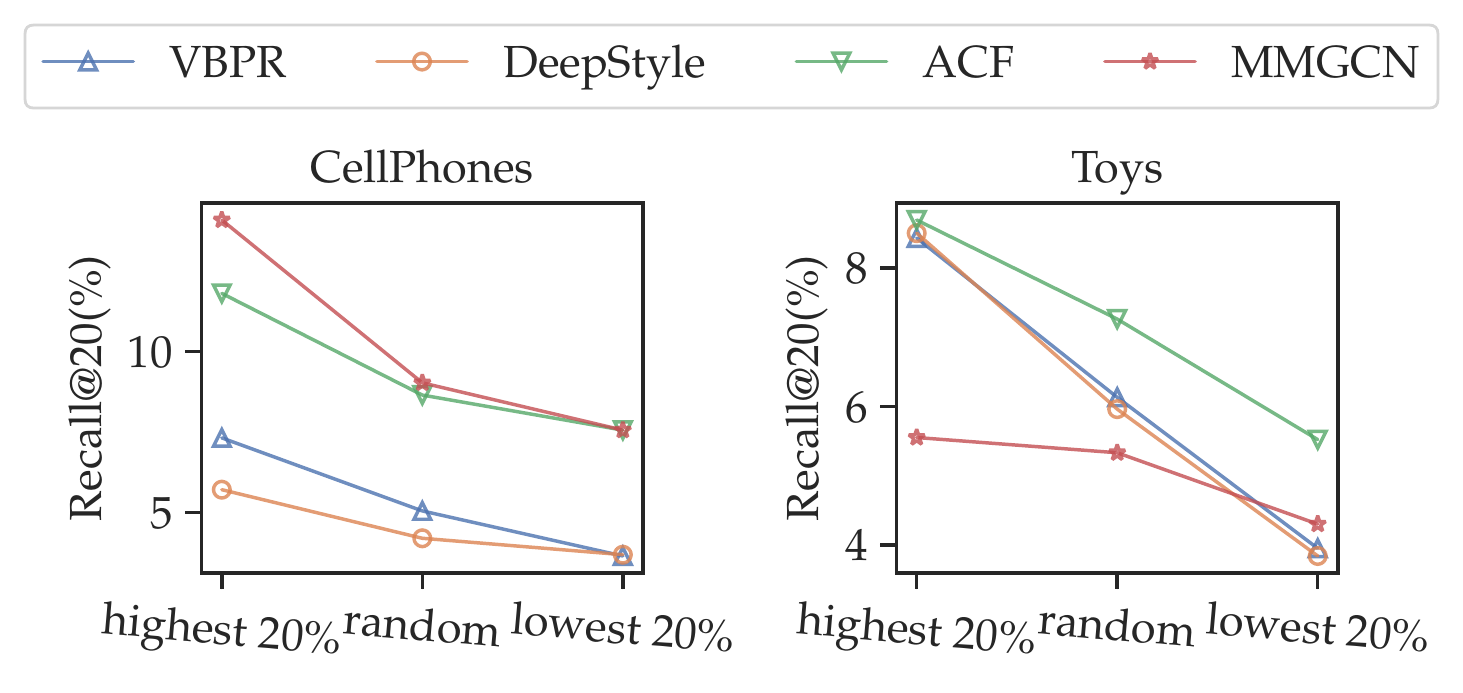}
    \caption{A pilot study on testing sets with different feature correlation degrees.}
    \label{fig:pilot}
\end{figure}
In this section, we first formulate the multimedia recommendation task. Then, to motivate our model design, we use a simple and intuitive experiments, to show that models are prone to be misled to learn the statistical spurious preference.
\subsection{Notations}
We denote the user set and item set as $\mathcal{U}$ and $\mathcal{I}$, respectively. $n=|\mathcal{I}|$ is the number of items. Each user $u \in \mathcal{U}$ has interacted with several items $\mathcal{I}^u$, which indicates that the preference score $y_{ui} = 1$ for $i \in \mathcal{I}^u$. Let $\bm x_u, \bm x_i \in \mathbb{R}^{d}$ denotes the input ID embedding of $u$ and $i$, respectively, where $d$ is the embedding dimension. 

Beside user-item interactions, each item is associated with multimodal content information. We denote the modality feature vector of item $i$ as $\bm{e}_i^m \in \mathbb{R}^{d_m}$ where $d_m$ denotes the dimension of the features and $m \in \mathcal{M}$ is the modality. For example, $m=v$ and $m=t$ denote the visual and textual modality, respectively. Our method is not fixed to the two modalities and multiple modalities can be involved. The purpose of multimedia recommendation is to accurately predict users' preferences $\hat{y}_{u i}$ by considering both user-item interactions and item multimodal content information.

\subsection{Multimedia Recommender Framework}
In this subsection, we introduce the general framework of multimedia recommender models. Our proposed method can be served as a flexible play-and-plug module for any multimedia recommender backbones. 
For multimedia recommender models, the calculation of preference score of user $u$ on item $i$ can be generalized as:
\begin{equation}
    \label{eq:score}
    \hat{y}_{u i} = f_\Theta(\bm{x}_u, \bm{x}_i, \bm{e}_{i}^m),  m \in \mathcal{M},
\end{equation}
where $f_\Theta(\cdot)$ represents different methods to model user-item interactions and item multimodal content information. $\Theta$ are the trainable parameters of the models. 

Specifically, since the feature vectors from different modalities usually have different dimensions, multimedia recommender models first need to transform raw modality features into a shared space:
\begin{equation}
    \label{eq:trs}
    \bar{\bm{e}}^m =  \bm{W}_m \bm{e}^m  + \bm{b}_m,
\end{equation}
where ${\bm{W}}_m \in \mathbb{R}^{d' \times d_m}$ and ${\bm{b}}_m \in \mathbb{R}^{d'}$ denote the trainable transformation matrix and bias vector, respectively. $d'$ is the dimension of the shared space. 

After that, taking VBPR \cite{He:2016ww} as an example, it simply conducts inner product between user and item representations:
\begin{equation}
    \hat{y}_{u i} = \bm{x}_u^\top \bm{x}_{i} + \sum_{m \in \mathcal{M}} {\bm{p}_u^m}^\top \bar{\bm{e}}_i^m,
\end{equation}
where $\bm{p}_u^m \in \mathbb{R}^{d'}$ is the trainable latent preference vector of $u$ towards attributes represented in modality $m$. For VBPR, trainable parameters $\Theta$ include the user and item ID embeddings $\bm{x}_u, \bm{x}_i$, trainable transformation parameters $\bm{W}_m, \bm{b}_m$ and user latent preference vectors $\bm{p}_u^m$.

Multimedia recommender models adopt Bayesian Personalized Ranking (BPR) loss \cite{Rendle:2009wp} to conduct the pair-wise ranking, which encourages the prediction of an observed entry to be higher than its unobserved counterparts:
\begin{equation}
    \label{eq:bpr}
    \mathcal{L}_{\text{BPR}}=-\sum_{(u,i,j) \in \mathcal{D}}  \ln \sigma\left(\hat{y}_{u i}-\hat{y}_{u j}\right),
\end{equation}
where $\mathcal{D} = \{ (u, i, j) | i \in \mathcal{I}_{u},  j \notin \mathcal{I}_{u} \}$ denotes the training set. $\mathcal{I}^{u}$ indicates the observed items associated with user $u$ and $(u, i, j)$ denotes the pairwise training triples where $i \in \mathcal{I}^u$ is the positive item and $j \notin \mathcal{I}^u$ is the negative item sampled from unobserved interactions. $\sigma(\cdot)$ is the sigmoid function.

\subsection{Pilot Study}
We first randomly select 80\% of historical interactions of each user to constitute the training set, 10\% for validation set and the remaining 10\% for testing set, which is consistent with previous work \cite{He:2016ww, Liu:2017ij, Chen:2017jj, Wei:2019hn}. Based on the randomly split testing set, we construct two additional testing sets with different data distributions. One only keeps 20\% of the items in the testing set with strongest correlation between modalities, and the other only keeps 20\% of the items with weakest correlation. The details of approximately estimating correlation degree are described in Section \ref{sec:eva}. Please note that only the data distribution of testing set shifts, and the training and validation set are the same as those in the first experiment.
\begin{figure*}
	\centering
    \includegraphics[width=0.9\linewidth]{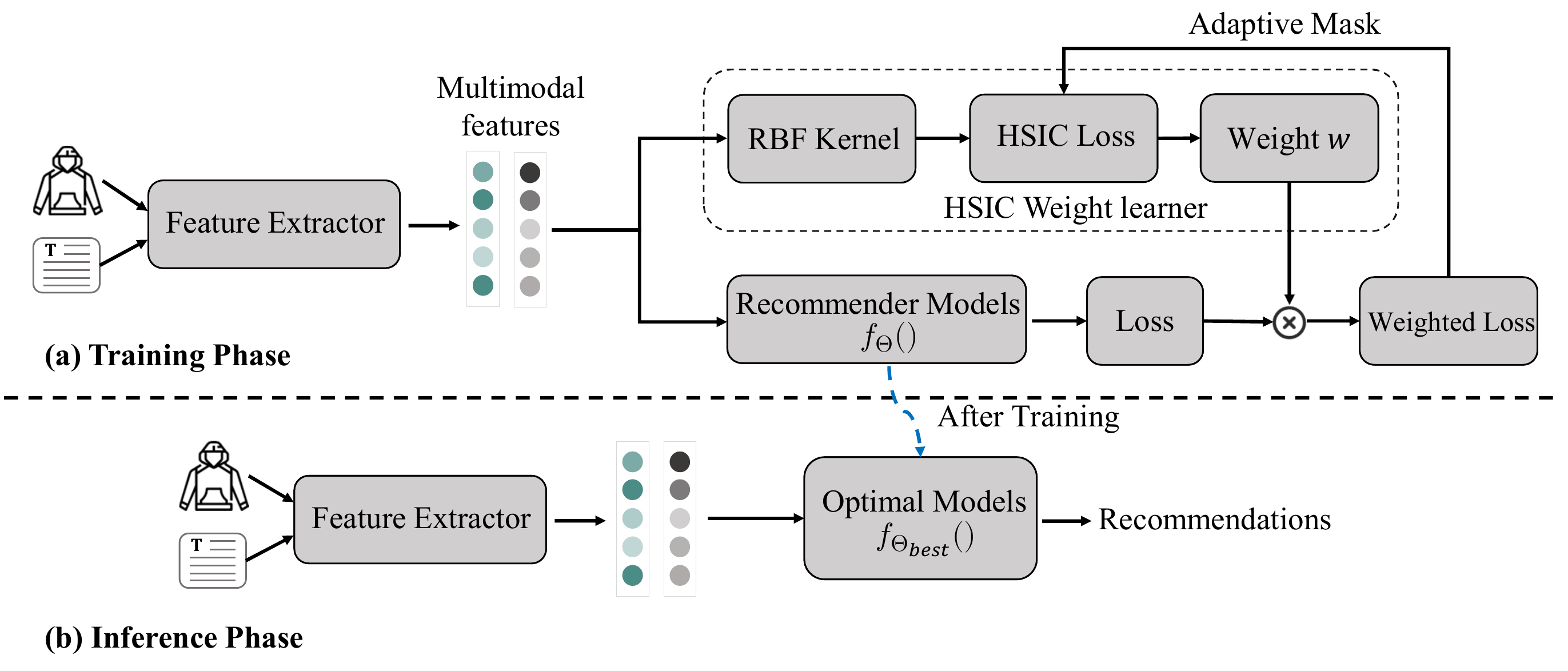}
    \caption{The overall framework of our proposed \themodel model, which could be served as a play-and-plug module for  multimedia recommendation backbones $f_{\Theta}()$. In the training phase, we alternately optimize model parameters and sample weights. In the inference phase, we can directly use the trained multimedia recommendation models to make recommendations.}
    \label{fig:model}
\end{figure*}

We perform four popular multimedia recommendation methods (VBPR \cite{He:2016ww}, DeepStyle \cite{Liu:2017ij}, ACF \cite{Chen:2017jj} and MMGCN \cite{Wei:2019hn}) on Amazon datasets (Cellphones and Toys) and the results are shown in Figure \ref{fig:pilot}, from which we can observe that backbones achieve much higher performances on the testing set with strong modal correlation while drop sharply on the testing set with weak modal correlation. The phenomenon indicates that models are prone to be misled to learn the spurious preference. On the one hand, the spurious preference still holds and even works better due to the strong modal correlation in testing set. On the other hand, however, when tested on the set with weak modal correlation, the spurious preference no longer holds and models fail to make stable predictions.

\section{The Proposed Method}
Motivated by the above analysis, in this section, we present our proposed modality decorrelation framework \themodel to guide backbone models to learn stable preferences. The overall framework is shown in Figure \ref{fig:model}. In the training phase, we alternately optimize (1) model parameters via minimizing weighted Bayesian Personalized Ranking (BPR) loss and (2) sample weights via minimizing task-relevant modal correlation (quantified through HSIC loss) in the weighted distribution. In the inference phase, we can directly use the trained multimedia recommendation models to make stable recommendations.

\subsection{Independence Testing Statistics}
\themodel aims to decorrelate the features of different modalities to learn stable preference. Firstly, the primary problem is how to measure the degree of dependence between any pair of modalities for each item. Since the feature vector is high-dimensional, it is infeasible to resort to histogram-based measures. In this paper, we adopt the kernel-based method Hilbert Schmidt Independence Criterion (HSIC) as independence testing measure \cite{gretton2005measuring, gretton2007kernel, greenfeld2020robust}. HSIC is the Hilbert-Schmidt norm of the cross-covariance operator between the distributions in Reproducing Kernel Hilbert Space (RKHS) \cite{fukumizu2009kernel}. In other words, HSIC is capable of evaluating the degree of correlation between two multi-dimensional and non-linear variables. Given two high-dimensional feature vectors $U \in \mathbb{R}^d$ and $V \in \mathbb{R}^d$, the formulation of HSIC is:
\begin{equation}
\begin{aligned}
\operatorname{HSIC}(U, V) &=\mathbb{E}_{u u^{\prime} v v^{\prime}}\left[k_{u}\left(u, u^{\prime}\right) k_{v}\left(v, v^{\prime}\right)\right] \\
&+\mathbb{E}_{u u^{\prime}}\left[k_{u}\left(u, u^{\prime}\right)\right] \mathbb{E}_{v v^{\prime}}\left[k_{v}\left(v, v^{\prime}\right)\right] \\
&-2 \mathbb{E}_{u v}\left[\mathbb{E}_{u^{\prime}}\left[k_{u}\left(u, u^{\prime}\right)\right] \mathbb{E}_{v^{\prime}}\left[k_{v}\left(v, v^{\prime}\right)\right]\right],
\end{aligned}
\end{equation}
where $u \in \mathbb{R}$ and $u^\prime \in \mathbb{R}$ represent the values in different dimensions of $U$. Similarly, $v \in \mathbb{R}$ and $v^\prime \in \mathbb{R}$ represent the values in different dimensions of $V$. $\mathbb{E}_{u u^{\prime} v v^{\prime}}$ denotes the expectation over pairs $(u, v)$ and $(u^\prime, v^\prime)$ drawn from $P(U, V)$. $k_u(\cdot)$ and $k_v(\cdot)$ are kernel functions. We utilize Radial Basis Function (RBF) kernel which is formulated as:
\begin{equation}
k(u, u^\prime)=\exp \left(-\frac{\|u-u^\prime\|_{2}^{2}}{\sigma^{2}}\right),
\end{equation}
where $\sigma$ is a free parameter. The empirical HSIC \cite{gretton2005measuring} is defined as:
\begin{equation}
\operatorname{HSIC}(U, V)=(d-1)^{-2} tr\left(\bm{K}_{U} \bm{P K}_{V} \bm{P}\right),
\end{equation}
where $tr(\cdot)$ denotes the trace of a matrix. $\bm{K}_{U} \in \mathbb{R}^{d \times d}$ and $\bm{K}_{V} \in \mathbb{R}^{d \times d}$ are the kernel matrices, that is, $\bm{K}_{U} := k(u_i, u_j)$ and $\bm{K}_{V} := k(v_i, v_j)$. $\bm{P} = \bm{I}_{d}-\frac{1}{d} \bm{1} \bm{1}_{d} \in \mathbb{R}^{d \times d}$ is the centering matrix, where $\bm{I}_{d}$ is the identity matrix and $\bm{1} \bm{1}_{d}$ is the matrix filled with ones. HSIC is zero if and only if the two random variables are independent:
\begin{equation}
    \operatorname{HSIC}(U, V) = 0 \Leftrightarrow U \perp V .
\end{equation}

\subsection{Adaptive Task-Relevant Mask}
\label{sec:mask}
Vanilla HSIC measures task-agnostic correlation since it essentially treats all feature dimensions equally. However, in multimedia recommendation task, different dimensions contribute differently since users usually pay different attention to various item attributes. As a result, vanilla HSIC might have difficulty in revealing the correlation of the factors which are important in recommendation task. To evaluate task-relevant correlation, we first define the task-relevant importance of dimension $i$ of modality $m$ $\alpha_i^m$. Inspired by previous work that investigates the gradients from the final predictions to the input features to visualize or interpret deep models \cite{DBLP:journals/corr/SimonyanVZ13, selvaraju2017grad, smilkov2017smoothgrad, lundberg2017unified}, we adopt gradient explanation to evaluate the contribution of different dimensions on the model output:
\begin{equation}
    \alpha_i^m = \sum_{j=1}^{d_m} \bigg |\frac{\partial \mathcal{L}_{BPR}}{\partial {W}_m} [i, j] \bigg |,
    \label{eq:grad-weight}
\end{equation}
where ${\bm{W}}_m \in \mathbb{R}^{d' \times d_m}$ is the trainable transformation matrix in Equation \ref{eq:trs}. Then, we normalize the task-relevant importance via softmax function:
\begin{equation}
    \bar{\alpha}_i^m =\frac{\operatorname{exp}({\alpha_i^m})}{\sum_{j=1}^{d'} \operatorname{exp}({\alpha_j^m})} \cdot d' . 
    \label{eq:grad-weight-norm}
\end{equation}
In this way, we can calculate the task-relevant HSIC between masked modal features $(\bm{\bar{\alpha}}^{m_1} \circ \bar{\bm{e}}^{m_1}, \bm{\bar{\alpha}}^{m_2} \circ \bar{\bm{e}}^{m_2})$, where $\bm{\bar{\alpha}}^m = (\bar{\alpha}_1^m, \cdots, \bar{\alpha}_{d'}^m) \in \mathbb{R}^{d'}$ and $\circ$ denotes element-wise multiplication.

\subsection{Modality Decorrelation with Re-weighting}
In this subsection, we introduce the proposed sample re-weighting framework which aims to estimate a weight for each item such that the features of different modalities in weighted distribution are decorrelated. We use $\bm{w} \in \mathbb{R}^n$ to denote the trainable sample weights where $n$ is the number of items. 

In the training phase, as shown in Algorithm \ref{algorithm}, we alternately optimize model parameters and sample weights. At each epoch, \themodel first fixes the sample weights and optimize the trainable parameters of multimedia recommendation models according to the weighted BPR loss. The sample weighting reassigns the importance of each sample to remove the statistical dependency between features.
\begin{equation}
\label{eq:theta}
    \Theta^{(q+1)} = \argmax_{\Theta} \sum_{(u,i,j) \in \mathcal{D}}  w^{(q)}_{i} \ln \sigma\left(\hat{y}_{u i}-\hat{y}_{u j}\right),
\end{equation}
where ${w}^{(q)}_{i} \in \mathbb{R}$ denotes the sample weight for item $i$ in $q$-th iteration and ${w}^{(0)}_{i}$ is initialized as $1$. $\hat{y}_{u i}$ is calculated through the backbone model in Equation \ref{eq:score}. Since negative item $j$ is random sampled, we only utilize the weight $w_i$ of positive item $i$. After that, \themodel finds optimal sample weights by minimizing the task-relevant HSIC between weighted features from different modalities while keeping the trainable parameters of multimedia recommendation models fixed:
\begin{equation}
\label{eq:weight}
\bm{w}^{(q+1)}=  \underset{\bm{w}}\argmin \sum_{i \in \mathcal{I}} \sum_{m_1} \sum_{m_2} \lambda \operatorname{HSIC}({w}^{(q)}_i \bm{\bar{\alpha}}^{m_1} \circ \bar{\bm{e}}^{m_1}_i, {w}^{(q)}_i \bm{\bar{\alpha}}^{m_2} \circ \bar{\bm{e}}^{m_2}_i),
\end{equation}
where $m_1, m_2 \in \mathcal{M}$ are two different modalities, $\bar{\bm{e}}^{m_1}_i, \bar{\bm{e}}^{m_2}_i \in \mathbb{R}^{d'}$ are the corresponding transformed modality feature vectors in Equation \ref{eq:trs}. $\bm{\bar{\alpha}}^{m_1}, \bm{\bar{\alpha}}^{m_2} \in \mathbb{R}^{d'}$ are the corresponding task-relevant importance in Equation \ref{eq:grad-weight-norm}. $\lambda$ is a hyper-parameter that balances the learning rate for updating model parameters $\Theta$ and sample weights $\bm{w}$.

To be noted, the sample weights are only used in the training phase to guide model training. In the inference phase, sample weights are not involved and only the trained backbone model is used to make recommendations.

\begin{algorithm}
\SetKwData{Left}{left}\SetKwData{This}{this}\SetKwData{Up}{up}
\SetKwFunction{Union}{Union}\SetKwFunction{FindCompress}{FindCompress}
\SetKwInOut{Input}{Input}\SetKwInOut{Output}{Output}
\caption{Training process of \themodel}
\label{algorithm}

\Input{Training dataset, maximum epoch $Epoch$, batch size $B$ and backbone model $f_{\Theta}(\cdot)$}
\Output{Optimal parameters $\Theta_{best}$ of backbone model.}
Initialize sample weights $\bm{w}^{(0)} \leftarrow \bm{1}^n$. \;
Initialize model parameters $\Theta^{(0)}$. \;
Initialize epoch indicator $q \leftarrow 1$. \;
Initialize the best iteration $q_{best} \leftarrow 1$. \;
\While{early stopping not reached and $q < Epoch$}
{
Keep $\bm{w}^{(q-1)}$ fixed and update model parameters $\Theta^{(q)}$ according to Equation \ref{eq:theta}. \;
Calculate task-relevant interpretations according to Equation \ref{eq:grad-weight} and \ref{eq:grad-weight-norm}. \;
Keep $\Theta^{(q)}$ fixed and update model parameters $\bm{w}^{(q)}$ according to Equation \ref{eq:weight}. \;
Update $q_{best}=q$, if better validation result achieved. \;
$q \leftarrow q + 1$. \;
} 
\Return{Optimal model parameters $\Theta^{q_{best}}$}
\end{algorithm}

\section{Experiments}
\begin{table}[]
    \caption{Statistics of the datasets}
    \begin{tabular}{ccccc}
    \toprule
    Dataset  & \#Users & \#Items & \#Interactions & Density \\
    \midrule
    Cellphones     & 27,879   & 10,429    & 165,192         & 0.00057 \\
    Beauty   & 22,363   & 12,101   & 172,188         & 0.00064 \\
    Baby     & 19,445   & 7,050    & 139,110         & 0.00101 \\
    Toys     & 19,412   & 11,924   & 145,004         & 0.00063 \\
    \bottomrule
    \end{tabular}
    \label{tab-dataset}
\end{table}

\begin{table*}
\caption{With random data split, performance comparison of our \themodel when plugged into different backbones in terms of Recall@20 (R@20), NDCG@20 (N@20) and Precision@20 (P@20). The best performance is highlighted \textbf{in bold}. $\Delta Improv.$ indicates relative improvements over InvRL in percentage. } 
\resizebox*{\textwidth}{!}{

\begin{tabular}{@{}ccccccccccccc@{}}
\toprule
\multirow{2.5}{*}{Model}                 & \multicolumn{3}{c}{Toys} & \multicolumn{3}{c}{CellPhones} & \multicolumn{3}{c}{Beauty}  & \multicolumn{3}{c}{Baby}     \\ 
\cmidrule(l){2-4} \cmidrule(l){5-7} \cmidrule(l){8-10} \cmidrule(l){11-13} 
                                      & R@20   & N@20   & P@20   & R@20     & N@20     & P@20     & R@20    & N@20    & P@20    & R@20    & N@20    & P@20     \\ \cmidrule(r){1-13}
VBPR                                   & 0.0613          & 0.0279          & 0.0032          & 0.0505          & 0.0217          & 0.0026          & 0.0598          & 0.0272          & 0.0033          & 0.0313          & 0.0140          & 0.0016          \\
+InvRL                                  & 0.0649          & 0.0298          & 0.0034          & 0.0504          & 0.0218          & 0.0025          & 0.0652          & 0.0304          & 0.0037          & \textbf{0.0337} & \textbf{0.0149} & \textbf{0.0019} \\
+\themodel                                   & \textbf{0.0669} & \textbf{0.0307} & \textbf{0.0035} & \textbf{0.0520} & \textbf{0.0223} & \textbf{0.0026} & \textbf{0.0721} & \textbf{0.0330} & \textbf{0.0041} & 0.0321          & 0.0140          & 0.0016          \\
\rowcolor{gray!20}
$\Delta Improv.$          & 3.08\%          & 3.02\%          & 2.94\%          & 3.17\%          & 2.29\%          & 4.00\%          & 10.58\%         & 8.55\%          & 10.81\%         & -         & -         & -        \\ \midrule
DeepStyle                              & 0.0596          & 0.0278          & 0.0032          & 0.0420          & 0.0183          & 0.0021          & 0.0626          & 0.0294          & 0.0035          & 0.0269          & 0.0117          & 0.0014          \\
+InvRL                                  & 0.0605          & 0.0277          & 0.0032          & 0.0484          & 0.0216          & 0.0025          & 0.0639          & 0.0298          & 0.0034          & 0.0288          & 0.0121          & 0.0015          \\
+\themodel                                   & \textbf{0.0648} & \textbf{0.0301} & \textbf{0.0035} & \textbf{0.0549} & \textbf{0.0253} & \textbf{0.0028} & \textbf{0.0739} & \textbf{0.0354} & \textbf{0.0042} & \textbf{0.0352} & \textbf{0.0158} & \textbf{0.0019} \\ 
\rowcolor{gray!20}
$\Delta Improv.$          & 7.11\%          & 8.66\%          & 9.38\%          & 13.43\%         & 17.13\%         & 12.00\%         & 15.65\%         & 18.79\%         & 23.53\%         & 22.22\%         & 30.58\%         & 26.67\%         \\ \midrule
ACF                                    & 0.0726          & \textbf{0.0385} & 0.0038          & 0.0865          & 0.0392          & 0.0044          & 0.0811          & 0.0397          & 0.0046          & 0.0488          & 0.0205          & 0.0025          \\
+InvRL                                  & 0.0727          & 0.0353          & 0.0038          & 0.0879          & 0.0396          & 0.0045          & 0.0815          & 0.0398          & 0.0046          & 0.0486          & 0.0205          & 0.0025          \\
+MODEST$ ^*$                            & \textbf{0.0746} & 0.0367          & \textbf{0.0039} & \textbf{0.0911} & \textbf{0.0410} & \textbf{0.0046} & \textbf{0.0837} & \textbf{0.0407} & \textbf{0.0047} & \textbf{0.0494} & \textbf{0.0210} & \textbf{0.0026} \\
\rowcolor{gray!20}
$\Delta Improv.$          & 2.61\%          & 3.97\%          & 2.63\%          & 3.64\%          & 3.54\%          & 2.22\%          & 2.70\%          & 2.26\%          & 2.17\%          & 1.65\%          & 2.44\%          & 4.00\%          \\ \midrule
MMGCN                                  & 0.0533          & 0.0231          & 0.0028          & 0.0903          & 0.0397          & 0.0046          & 0.0699          & 0.0311          & 0.0040          & 0.0586          & 0.0248          & 0.0031          \\
+InvRL                                  & 0.0539          & 0.0233          & 0.0028          & 0.0911          & 0.0398          & 0.0044          & 0.0734          & 0.0337          & 0.0042          & 0.0565          & 0.0241          & 0.0030          \\
+\themodel                                   & \textbf{0.0549} & \textbf{0.0238} & \textbf{0.0029} & \textbf{0.0942} & \textbf{0.0421} & \textbf{0.0048} & \textbf{0.0788} & \textbf{0.0349} & \textbf{0.0044} & \textbf{0.0616} & \textbf{0.0263} & \textbf{0.0033} \\
\rowcolor{gray!20}
$\Delta Improv.$          & 1.86\%          & 2.15\%          & 3.57\%          & 3.40\%          & 5.78\%          & 9.09\%          & 7.36\%          & 3.56\%          & 4.76\%          & 9.03\%          & 9.13\%          & 10.00\%         \\ \midrule
Averaged $\Delta Improv.$ & 3.66\%          & 4.45\%          & 4.63\%          & 5.91\%          & 7.18\%          & 6.83\%          & 9.07\%          & 8.29\%          & 10.32\%         & 7.04\%          & 9.03\%          & 6.22\%          \\ \bottomrule
\end{tabular}
}
\label{tab-main}
\end{table*}

In this section, we conduct experiments on four widely used real-world datasets. We firstly describe the experimental settings, including datasets, baselines, evaluation and implementation details. Then, we report and discuss the experimental results to answer the following research questions:
\begin{itemize}
    \item \textbf{RQ1:} How does our method perform when plugged into different multimedia recommendation methods?
    \item \textbf{RQ2:} Does our method achieve stable recommendation performance when the data distribution of testing set is different from that of training set?
    \item \textbf{RQ3:} How sensitive is our model with different hyper-parameter settings?
\end{itemize}
\subsection{Experiments Settings}
\subsubsection{Datasets}
We conduct experiments on four categories of widely used Amazon review dataset introduced by \citet{McAuley:2015ip}: `Beauty', `Baby', 'Toys and Games', 'Cell Phones and Accessories', which are named as \textbf{Beauty}, \textbf{Baby}, \textbf{Toys} and \textbf{Cellphones} in brief. We use the 5-core version of Amazon datasets where each user and item have 5 interactions at least. The statistics of these datasets are summarized in Table \ref{tab-dataset}. Amazon dataset includes visual modality and textual modality. The 4,096 dimensional visual features have been extracted and published\footnote{\url{http://jmcauley.ucsd.edu/data/amazon/links.html}}. We also extract sentence embeddings as textual features by concatenating the title, descriptions, categories, and brand of each item and then utilize sentence-transformers \cite{Reimers:2019iz} to obtain 1,024 dimensional sentence embeddings. 
\subsubsection{Baseline Models}
To evaluate the effectiveness of our proposed model, we plug it into several state-of-the-art recommendation models.
\begin{itemize}
    \item \textbf{VBPR} \cite{He:2016ww}: Based upon the BPR model, it integrates the visual features and ID embeddings of each item as its representation and feed them into Bayesian Personalized Ranking (BPR) framework. In our experiments, we concatenate multi-modal features as the content information to predict the interactions between users and items.
    \item \textbf{DeepStyle} \cite{Liu:2017ij} proposes that style information indicates the preferences of users which has significant effect in recommendation. It derives user preferences for different styles from visual contents.
    \item \textbf{ACF} \cite{Chen:2017jj} introduces an item-level and component-level attention model which learns to select informative components of multimedia items and therefore better understand the user preferences for item contents.
    \item \textbf{MMGCN} \cite{Wei:2019hn} is one of the state-of-the-art multimodal recommendation methods, which constructs modal-specific graphs and refines modal-specific representations for users and items. It aggregates all model-specific representations to obtain the representations of users or items for prediction.
\end{itemize}
In addition, invariant risk minimization (IRM) \cite{arjovsky2019invariant, wang2022invariant} based methods aim to learn invariant representations across various environments. We also conduct experiments on IRM-based method \textbf{InvRL} \cite{du2022invariant} for comparison.
\begin{table*}
\caption{With data split where training and testing set have different data distribution, performance comparison of our \themodel when plugged into different backbones in terms of Recall@20 (R@20), NDCG@20 (N@20) and Precision@20 (P@20). The best performance is highlighted \textbf{in bold}. $\Delta Improv.$ indicates relative improvements over InvRL in percentage. }
\resizebox*{\textwidth}{!}
{
\begin{tabular}{@{}ccccccccccccc@{}}
\toprule
\multirow{2.5}{*}{Model}                 & \multicolumn{3}{c}{Toys} & \multicolumn{3}{c}{CellPhones} & \multicolumn{3}{c}{Beauty}  & \multicolumn{3}{c}{Baby}     \\ 
\cmidrule(l){2-4} \cmidrule(l){5-7} \cmidrule(l){8-10} \cmidrule(l){11-13} 
                                      & R@20   & N@20   & P@20   & R@20     & N@20     & P@20     & R@20    & N@20    & P@20    & R@20    & N@20    & P@20     \\ \cmidrule(r){1-13}
VBPR                                   & 0.0395          & 0.0168          & 0.0020          & 0.0365          & 0.0162          & 0.0018          & 0.0516          & 0.0243          & 0.0027          & 0.0311          & 0.0151          & 0.0016          \\
+InvRL                                  & 0.0421          & 0.0181          & 0.0022          & 0.0361          & 0.0161          & 0.0018          & 0.0562          & 0.0234          & 0.0029          & \textbf{0.0342} & \textbf{0.0199} & \textbf{0.0018} \\
+\themodel                                   & \textbf{0.0455} & \textbf{0.0200} & \textbf{0.0024} & \textbf{0.0383} & \textbf{0.0165} & \textbf{0.0019} & \textbf{0.0639} & \textbf{0.0278} & \textbf{0.0034} & 0.0338          & 0.0188          & 0.0017          \\
\rowcolor{gray!20}
$\Delta Improv.$          & 8.08\%          & 10.50\%         & 9.09\%          & 6.09\%          & 2.48\%          & 5.56\%          & 13.70\%         & 18.80\%         & 17.24\%         & -               & -               & -               \\ \midrule
DeepStyle                              & 0.0384          & 0.0178          & 0.0020          & 0.0369          & 0.0165          & 0.0015          & 0.0513          & 0.0230          & 0.0027          & 0.0212          & 0.0093          & 0.0011          \\
+InvRL                                  & 0.0403          & 0.0181          & 0.0020          & 0.0370          & 0.0166          & 0.0018          & 0.0519          & 0.0232          & 0.0027          & 0.0232          & 0.0099          & 0.0012          \\
+\themodel                                   & \textbf{0.0441} & \textbf{0.0194} & \textbf{0.0023} & \textbf{0.0422} & \textbf{0.0190} & \textbf{0.0021} & \textbf{0.0699} & \textbf{0.0319} & \textbf{0.0037} & \textbf{0.0305} & \textbf{0.0130} & \textbf{0.0016} \\
\rowcolor{gray!20}
$\Delta Improv.$          & 9.43\%          & 7.18\%          & 15.00\%         & 14.05\%         & 14.46\%         & 16.67\%         & 34.68\%         & 37.50\%         & 37.04\%         & 31.47\%         & 31.31\%         & 33.33\%         \\ \midrule
ACF                                    & 0.0552          & \textbf{0.0252} & 0.0028          & 0.0750          & 0.0348          & 0.0038          & 0.0904          & 0.0383          & 0.0047          & 0.0326          & 0.0118          & 0.0017          \\
+InvRL                                  & 0.0536          & 0.0238          & 0.0028          & 0.0729          & 0.0327          & 0.0037          & 0.0924          & 0.0410          & 0.0048          & 0.0359          & 0.0131          & 0.0018          \\
+\themodel                                   & \textbf{0.0564} & 0.0244          & \textbf{0.0029} & \textbf{0.0751} & \textbf{0.0389} & \textbf{0.0039} & \textbf{0.0980} & \textbf{0.0473} & \textbf{0.0051} & \textbf{0.0481} & \textbf{0.0188} & \textbf{0.0024} \\
\rowcolor{gray!20}
$\Delta Improv.$          & 5.22\%          & 2.52\%          & 3.57\%          & 3.02\%          & 18.96\%         & 5.41\%          & 6.06\%          & 15.37\%         & 6.25\%          & 33.98\%         & 43.51\%         & 33.33\%         \\ \midrule
MMGCN                                  & 0.0430          & 0.0170          & 0.0022          & 0.0756          & 0.0389          & 0.0038          & 0.0640          & 0.0246          & 0.0034          & 0.0637          & 0.0276          & 0.0030          \\
+InvRL                                  & 0.0438          & 0.0187          & 0.0023          & 0.0760          & 0.0391          & 0.0038          & 0.0685          & 0.0267          & 0.0036          & 0.0641          & 0.0277          & 0.0031          \\
+\themodel                                   & \textbf{0.0456} & \textbf{0.0199} & \textbf{0.0024} & \textbf{0.0816} & \textbf{0.0468} & \textbf{0.0041} & \textbf{0.0737} & \textbf{0.0277} & \textbf{0.0039} & \textbf{0.0702} & \textbf{0.0347} & \textbf{0.0033} \\
\rowcolor{gray!20}
$\Delta Improv.$          & 4.11\%          & 6.42\%          & 4.35\%          & 7.37\%          & 19.69\%         & 7.89\%          & 7.59\%          & 3.75\%          & 8.33\%          & 9.52\%          & 25.27\%         & 6.45\%          \\ \midrule
Averaged $\Delta Improv.$ & 6.71\%          & 6.65\%          & 8.00\%          & 7.63\%          & 13.90\%         & 8.88\%          & 15.51\%         & 18.85\%         & 17.22\%         & 18.45\%         & 23.64\%         & 16.89\%         \\ \bottomrule
\end{tabular}
}
\label{tab-ood}
\end{table*}

\subsubsection{Evaluation}
\label{sec:eva}
We conduct two experiments to validate the rationality of our proposed method:

In the first experiment, we aim to evaluate the performance of our method when training, validation and testing set have similar data distribution by random spilt. For each dataset, we randomly select 80\% of historical interactions of each user to constitute the training set, 10\% for validation set and the remaining 10\% for testing set.

In the last two experiments, we aim to evaluate the out-of-distribution (OOD) performance when the training set and testing set have different data distributions. It is more challenging since the spurious dependency between true preference of users and unimportant modalities of items learned in training set might not work in testing set. We denote the co-occurrence joint distribution between textual and visual modality as $P(\bm{e}^v, \bm{e}^{t})$. Since there are no related methods to construct the OOD dataset of modal correlation, we propose to approximately let $P_{train}(\bm{e}^v, \bm{e}^t) \neq P_{test}(\bm{e}^v, \bm{e}^t)$ using two different methods:
\begin{itemize}
    \item The first method is to approximately estimate $P(\bm{e}^v, \bm{e}^{t})$ directly. Inspired by Permutation Weighting \cite{arbour2021permutation} which estimates propensity scores using binary classifier-based estimation, we construct the estimator with a MLP binary classifier, of which the input is multimodal feature pair $(\bm{e}^v, \bm{e}^t)$  and the output is the probability of whether the feature pair match. During training the estimator, we set $(\bm{e}_i^v, \bm{e}_i^t)$ as positive pairs and randomly sample $(\bm{e}_i^v, \bm{e}_j^t)_{i \neq j}$ as negative pairs. Secondly, after training, we feed the estimator with $(\bm{e}_i^v, \bm{e}_i^t)$ and the output denotes the probability value of $p(\bm{e}_i^v, \bm{e}_i^t)$. We select 20\% items with lowest probabilities and only keep the interactions including them in the testing set. Please note that only the data distribution of testing set shifts, and the training and validation set are the same as those in the first experiment. The performances are reported in Table \ref{tab-ood}.
    \item Additionally, we assume that different datasets have different distribution $P(\bm{e}^v, \bm{e}^{t})$ and create synthetic datasets by mixing two different datasets. For example, we use 80\% interactions from Baby and 10\% interactions from Toys to build training set, 10\% interactions from Baby and 10\% interactions from Toys to build validation set, 10\% interactions from Baby and 80\% interactions from Toys to build testing set. In this way, training and testing set have different mixing ratio and thus $P_{train}(\bm{e}^v, \bm{e}^t) \neq P_{test}(\bm{e}^v, \bm{e}^t)$. We denote the synthetic dataset as \textbf{8:1:1 from Baby + 1:1:8 from Toys}. Similarly, we use different datasets and different mixing ratios to build multiple IID and OOD datasets. The performances are reported in Table \ref{tab-mix}.
\end{itemize}
Please note that the sample weights are only trained with training set and only used in training to guide models. They are not used in testing stage or splitting OOD testing set.

\begin{table*}[]
\resizebox*{\textwidth}{!}{
\begin{tabular}{@{}ccccccccccccc@{}}
\toprule
\multirow{2}{*}{Dataset}                             & \multicolumn{3}{c}{VBPR}    & \multicolumn{3}{c}{DeepStyle} & \multicolumn{3}{c}{ACF}     & \multicolumn{3}{c}{MMGCN}   \\ \cmidrule(l){2-4} \cmidrule(l){5-7} \cmidrule(l){8-10} \cmidrule(l){11-13} 
                                                     & Backbone & +\themodel   & $\Delta Improv.$  & Backbone  & +\themodel    & $\Delta Improv.$  & Backbone & +\themodel   & $\Delta Improv.$  & Backbone & +\themodel   & $\Delta Improv.$  \\ \midrule
8:1:1 from Baby   + 8:1:1 from Toys (IID)        & 0.0437   & 0.0450 & 2.97\%  & 0.0423    & 0.0434  & 2.60\%  & 0.0567   & 0.0580 & 2.29\%  & 0.0489   & 0.0497 & 1.53\%  \\
8:1:1 from Baby   + 1:1:8 from Toys (OOD)            & 0.0268   & 0.0284 & 5.97\%  & 0.0311    & 0.0331  & 6.43\%  & 0.0267   & 0.0285 & 6.74\%  & 0.0302   & 0.0329 & 8.94\%  \\
8:1:1 from Baby   + 2.5:2.5:5 from Toys (OOD)        & 0.0324   & 0.0357 & 10.19\% & 0.0341    & 0.0379  & 11.14\% & 0.1229   & 0.1314 & 6.92\%  & 0.0432   & 0.0472 & 9.26\%  \\ \midrule
8:1:1 from Beauty   + 8:1:1 from Toys (IID)      & 0.0576   & 0.0581 & 0.87\%  & 0.0566    & 0.0603  & 6.54\%  & 0.0745   & 0.0765 & 2.68\%  & 0.0604   & 0.0619 & 2.48\%  \\
8:1:1 from Beauty   + 1:1:8 from Toys (OOD)          & 0.0391   & 0.0431 & 10.23\% & 0.0431    & 0.0497  & 15.31\% & 0.0409   & 0.0450  & 10.02\% & 0.0416   & 0.0438 & 5.29\%  \\
8:1:1 from Beauty   + 2.5:2.5:5 from Toys (OOD)  & 0.0403   & 0.0466 & 15.63\% & 0.0484    & 0.0518  & 7.02\%  & 0.0461   & 0.0472 & 2.39\%  & 0.0384   & 0.0435 & 13.28\% \\ \bottomrule
\end{tabular}
}
\caption{Performance comparison (Recall@20) on the mixed datasets. $\Delta Improv.$ indicates relative improvements over backbones in percentage.}
\label{tab-mix}
\end{table*}

\subsubsection{Implementation details}
We implemented our method in PyTorch and the embedding dimension $d$ is fixed to 64 for all models to ensure fair comparison. We optimize all models with the Adam \cite{Kingma:2015us} optimizer, where the batch size is fixed at 1024. We use the Xavier initializer \cite{Glorot:2010uc} to initialize the model parameters. The optimal hyper-parameters were determined via grid search on the validation set: the learning rate is tuned amongst \{0.0001, 0.0005, 0.001, 0.005\}, the coefficient of $L_2$ normalization is searched in \{1e-5, 1e-4, 1e-3, 1e-2\}, and the dropout ratio in \{0.0, 0.1, $\cdots$, 0.8\}. Besides, we stop training if recall@20 on the validation set does not increase for 10 successive epochs. 

\subsection{Overall Performance Comparison (RQ1)}
We start by conducting experiments on randomly split dataset, where samples in training, validation and testing sets share relative similar data distribution. The overall performances of \themodel are reported in Table 2, from which we have the following observations:
\begin{itemize}
    \item Our proposed method can provide significant improvements, verifying the effectiveness of our method. Specifically, our method improves over the backbone multimedia recommendation models by 8.8\%, 22.1\%, 3.1\%, 6.3\%, and improves IRM-based method InvRL 3.1\%, 14.6\%, 2.7\%, 5.4\% in terms of Recall@20 on average on VBPR, DeepStyle, ACF and MMGCN, respectively.  This indicates that our proposed method is well-designed for multimedia recommendation by decorrelating different modalities and thus the backbone models are guided to focus on the modalities that can reveal users' true preferences instead of unimportant modalities.
    \item Compared with VBPR and DeepStyle, ACF and MMGCN achieve better performance overall, which indicates that attention mechanism and graph neural networks can help multimedia recommendation models to model the complicated interactions between users and multimodal items. Even plugged into these powerful models, our method can achieve consistent improvement.
\end{itemize}
\subsection{OOD Performance Comparison (RQ2)}
In this subsection, we conduct performance comparison on the data split where the training set and testing set have different data distributions, to investigate whether our proposed method can provide stable predictions when data distribution shifts.

Firstly, we construct OOD testing set by approximately estimating $P(\bm{e}^v, \bm{e}^{t})$. The performances are reported in Table \ref{tab-ood} and Table \ref{tab-mix}, from which we have the following observations:
\begin{itemize}
    \item Compared with the performances in Table \ref{tab-main}, with the same training and validation set, when testing distribution shifts, the performances of backbones drop sharply, indicating that backbone models are prone to be misled by the strong statistical correlation between modalities, thus the true causal effect of each modality cannot be accurately estimated by models. The spurious preference towards unimportant modalities learned in the training set cannot guarantee to be as effective in the testing set due to the distribution shift. 
    \item \themodel has relatively greater improvements against backbones and InvRL than the results with random data split, verifying the effectiveness of our method in making stable recommendation when data distribution shifts. Specifically, our method improves over the backbone multimedia recommendation models by 13.2\%, 27.2\%, 14.6\%, 9.8\%, and improves InvRL 6.7\%, 22.4\%, 12.1\%, 7.2\% in terms of Recall@20 on average on VBPR, DeepStyle, ACF and MMGCN, respectively. 
\end{itemize}

We also construct synthetic datasets by mixing different datasets. The performances are reported in Table \ref{tab-mix}, from which we have the following observations:
\begin{itemize}
    \item \themodel can provide significant improvements on both IID datasets and OOD datasets. Specifically, it gains 2.75\% and 9.05\% average improvements on IID datasets and OOD datasets when plugged into the four backbones in terms of Recall@20, respectively. 
    \item \themodel gains much greater improvements on OOD datasets, demonstrating our claim that the proposed method can discover stable preference reflecting true causal effect of each modalities, which is free of data biases introduced by the distribution shifts between training and testing set.
\end{itemize}

\subsection{In-depth Analysis (RQ3)}
\begin{figure}[t]
	\centering
	\includegraphics[width=\linewidth]{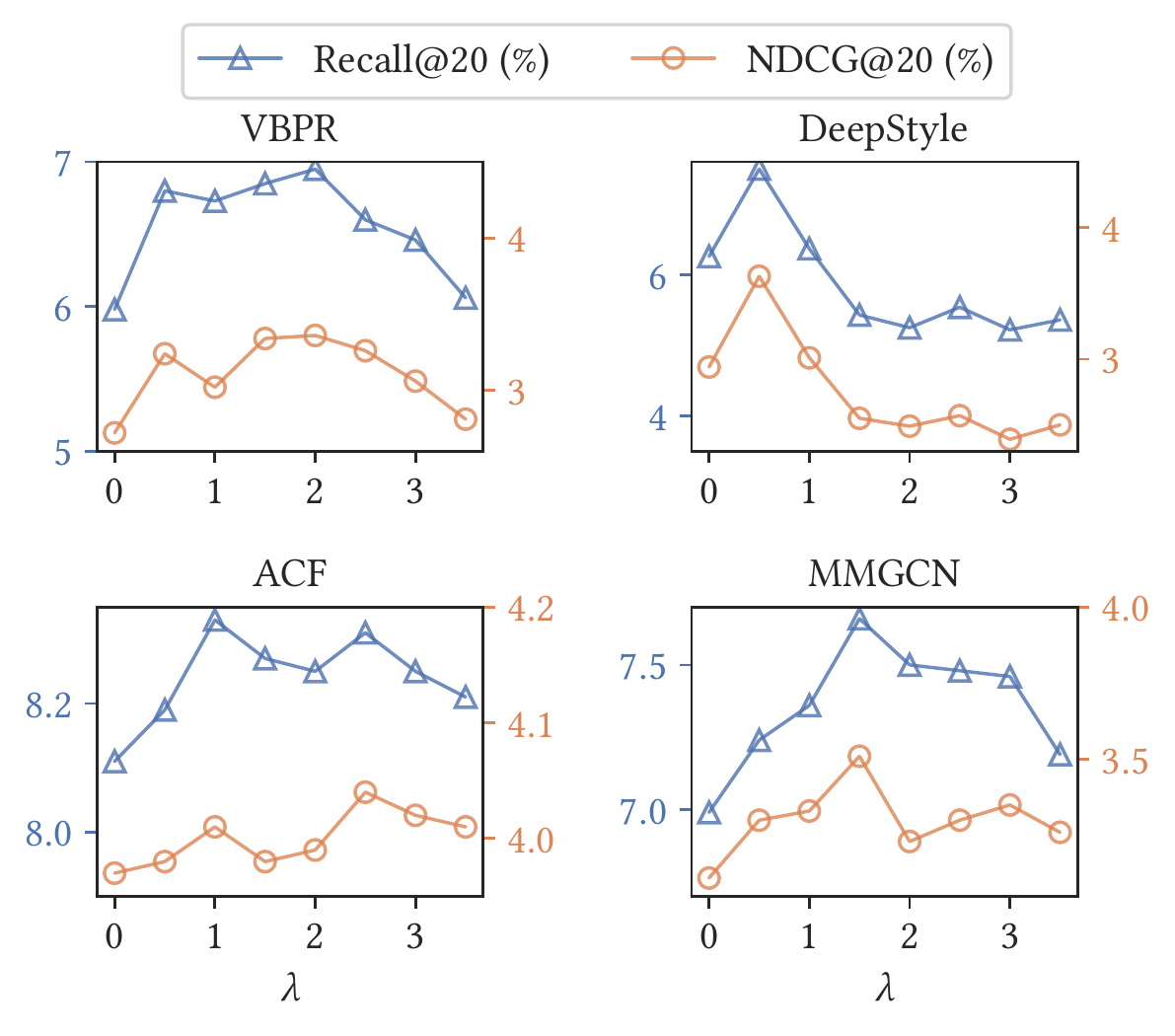}
    \caption{Performances comparison over varied \(\lambda\) for different backbones on Beauty.}
    \label{fig:hyperparams}
\end{figure}

We conduct sensitive analysis with different coefficient $\lambda$ in Equation \ref{eq:weight} on datasets under the random splitting. $\lambda = 0$ means our proposed method is not included and the model is degenerated to the vanilla backbone. Figure \ref{fig:hyperparams} reports the performance comparison, from which we can observe that our method gains improvement between $\lambda=0$ and $\lambda=0.5$, which validates the rationality of our proposed method. Even with a small learning rate for updating sample weights, the backbone models can be forced to focus on the modalities that can reveal users' true preferences, and thus boost the recommendation performance. Furthermore, the performances first grow as $\lambda$ becomes larger, since the sample weights are optimized and converged by a lager learning rate. The trend, however, fluctuates when $\lambda$ continues to increase. A too-large $\lambda$ will cause a sharp drop in recommendation performance. Notably, the majority of performances exceed the backbones, proving the effectiveness of our method.

\subsection{Visualization}

\begin{figure}[t]
    \centering
    \includegraphics[width=0.8\linewidth]{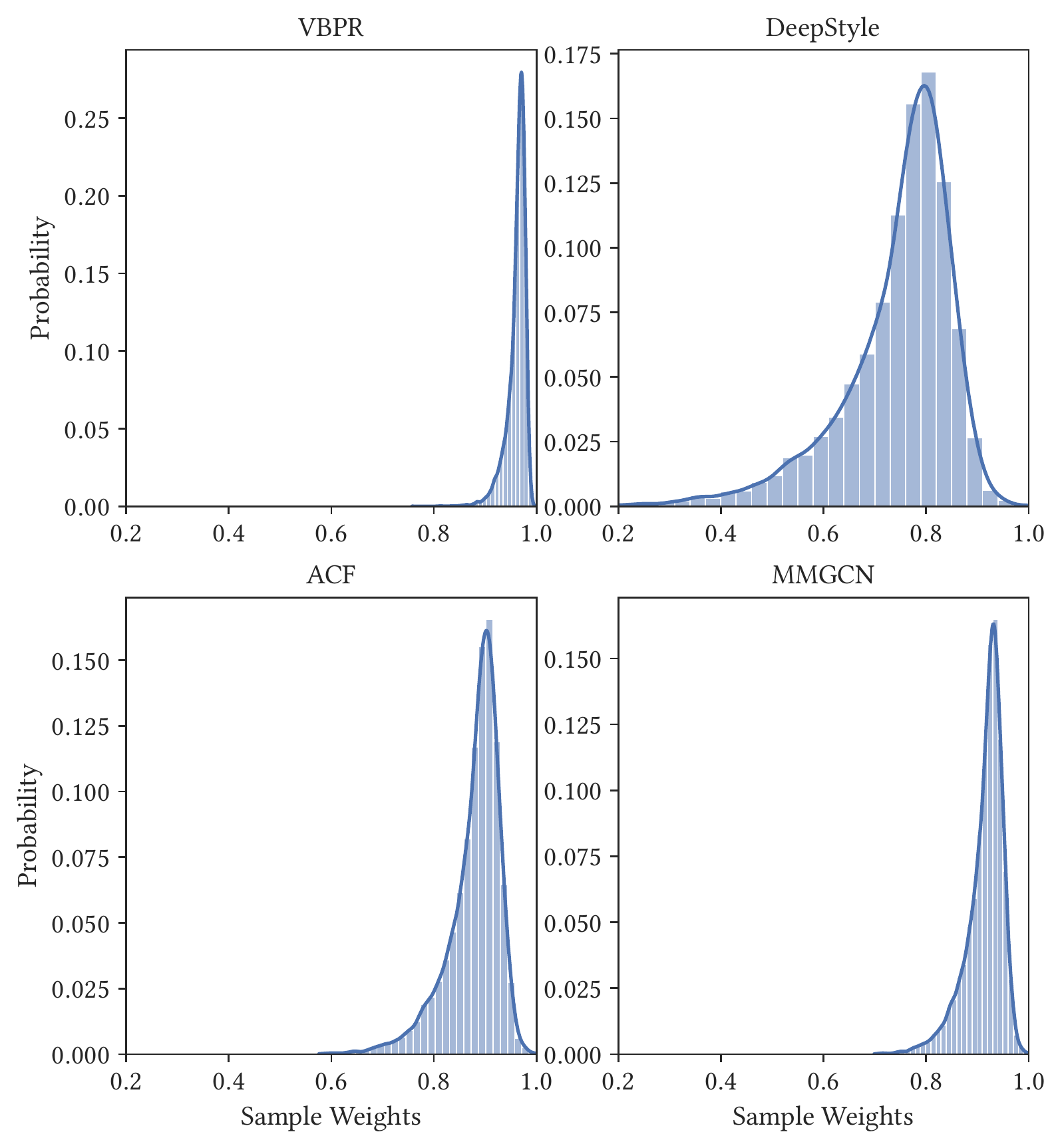}
    \caption{The histogram of sample weights for different backbones on Beauty.}
    \label{fig:histogram}
\end{figure}

In Figure \ref{fig:histogram}, we visualize the probability distribution of sample weights for the Beauty dataset. The value near 1.0 means no change of sample weight while the value near 0.0 means a sharp change of sample weight in the loss function. We can observe that the sample weights are around $0.8$ to $1.0$ for VBPR and MMGCN, $0.4$ to $1.0$ for DeepStyle and $0.6$ to $1$ for ACF. Although the means are different, the distributions of different models are roughly gamma-distributed. By marginalizing the samples of which the modalities are strongly correlated, models can avoid being misled by the correlations between modalities and focus on the modalities which make more contribution to purchase decision.

\section{Related Work}
\subsection{Multimedia Recommendation}
Traditional work on multimedia recommendation \cite{deldjoo2020recommender, He:2016ww, Kang:2017ds, Liu:2017ij, Chen:2017jj, cui2018mv, xun2021we,tan2021counterfactual} extends the vanilla CF framework by incorporating multimodal contents as side information in addition to item representations. For example, VBPR \cite{He:2016ww} extended the framework of Bayesian Personalized Ranking (BPR) by concatenating content vectors with item embeddings to improve the performance.  DeepStyle \cite{Liu:2017ij} derives user preferences for different styles from visual contents. ACF \cite{Chen:2017jj} introduces an item-level and component-level attention model to better understand the user preferences for item contents. Recently, due to its effectiveness in modeling graph-structure data \cite{Kipf:2017tc, Velickovic:2018we, Wang:2019er, He:2020gd, Wu:2019ke, zhang2022dynamic}, Graph Neural Networks (GNNs) have been introduced into multimedia recommendation systems \cite{Wei:2019hn,Wei:2020ko,li2020hierarchical,zhang_lattice, lu2021multi, liu2021concept,kim2022mario}. For example, MMGCN \cite{Wei:2019hn} constructed modal-specific graph and refine modal-specific representations for users and items. 
GRCN \cite{Wei:2020ko} refined user-item interaction graph by identifying the false-positive feedback and prunes the corresponding noisy edges in the interaction graph. LATTICE \cite{zhang_lattice} proposes to explicitly model item relationships, including collaborative relationships and semantic relationships, and utilizes graph structure learning \cite{Zhu:2021ue} to build semantic item graph.  Additionally, there are increasing research interests on introducing contrastive learning framework into multimedia recommendation systems\cite{zhang2022latent, wei2021contrastive, zhang2022ccl4rec}. \citet{wei2021contrastive} aim to maximize the mutual information between item content and collaborative signals to alleviate the cold-start problem. MICRO \cite{zhang2022latent} employs contrastive learning to conduct fine-grained multimodal fusion.

\subsection{Stable Learning}
Since the correlation between features could mislead model training and thus affect model predictions, several methods have been proposed to eliminate this correlation.  Inverse Propensity Weighting (IPW) approaches \cite{arbour2021permutation, schnabel2016recommendations, wang2022unbiased} can remove confounding but propensity scores are hard to accurately estimate with high variance. Recently, stable learning algorithms \cite{kuang2018stable, cui2022stable} have shown effectiveness to perform feature decorrelation. Stable learning aims to learn a stable predictive model that achieves uniformly good performance on any unknown test data distribution \cite{shen2021towards}. The framework of these stable learning algorithms usually consists of two steps, namely importance sampling and weighted training \cite{xu2021stable}. In the importance sampling step, the stable learning algorithms aim to learn a weight for each sample, such that the features in weighted distribution are statistical independent. And then weighted training is conducted to train models on weighted feature distribution. Various decorrelation methods have been proposed to learn samples weights in linear \cite{shen2020stable,kuang2020stable} or non-linear deep models \cite{zhang2021deep, fan2021generalizing, luo2022deep, liu2023deep}. For example, \citet{kuang2020stable} propose to learn a set of sample weights such that the weighted distribution of treatment and confounder could be independent. Sample Reweighted Decorrelation Operator (SRDO) \cite{shen2020stable} generates some unobserved samples, and trains a binary classifier to get the probabilities of observation for weighting the observed samples. StableNet \cite{zhang2021deep} proposes a non-linear decorrelation method which adopts Random Fourier Features. The sample weights and representation learning model are optimized alternatively.  StableGNN \cite{fan2021generalizing} proposes to decorrelate features in graph neural networks. \citet{zhang2021stable} propose to permits both locally and globally stable learning and prediction on graphs.  
\section{Conclusion}
In this paper, we have proposed \themodel which aims to avoid models being misled by the spurious preference towards unimportant modalities, which could be easily served as a play-and-plug module for existing multimedia recommendation models. Specifically, \themodel utilizes sample re-weighting techniques such that the features of different modalities in weighted distribution are decorrelated. Extensive experiments on four real-world datasets and four state-of-the-art multimedia recommendation backbones have been conducted to demonstrate that \themodel achieves superior performance on public benchmarks and the synthetic OOD datasets.

\begin{acks}
This work is supported by National Key Research and Development Program (2019QY1601, 2019QY1600) and  National Natural Science Foundation of China (62141608, 62206291, 62236010).
\end{acks}

\bibliographystyle{ACM-Reference-Format}
\bibliography{sigir2023}

\end{document}